# R-CA: A Routing-based Dynamic Channel Assignment Algorithm for Wireless Mesh Networks

Weifeng Sun, Rong Cong, Feng Xia, Xiao Chen, Zhenquan Qin*
School of Software, Dalian University of Technology, Dalian 116620, China
e-mail: wfsun@dlut.edu.cn, cong4218@163.com, f.xia@ieee.org, cx-king@126.com, qzq@dlut.edu.cn

*Abstract*—Even though channel assignment has been studied for years, the performance of most IEEE 802.11-based multi-hop wireless networks such as wireless sensor network (WSN), wireless mesh network (WMN), mobile ad hoc network (MANET) is limited by channel interference. Properly assigning orthogonal channels to wireless links can improve the throughput of multi-hop networks. To solve the dynamic channel assignment problem, a routing-based channel assignment algorithm called R-CA is proposed. R-CA can allocate channels for wireless nodes when needed and free channels after data transmission. Thus more channel resource can be explored by wireless nodes. Simulation results show that R-CA can effectively enhance the network throughput and packet delivery rate.

I. INTRODUCTION

Wireless mesh network (WMN), separated from the mobile ad hoc networks (MANETs), is emerging as a promising technology for low-cost, high-capacity, flexibility and ubiquitous broadband Internet access. It integrates the advantages of WLAN and ad hoc networks. In a wireless mesh network, a collection of wireless access routers provide connectivity to mobile clients acting as access points in a traditional wireless LAN, but access routers communicate with each other wirelessly, potentially over multiple hops. A small fraction of those access routers are wired to the Internet and serve as Internet gateways for the rest of the network. Wireless mesh networks based on commodity 802.11 hardware and employing self-configuring ad hoc networking techniques can offer wider coverage with less expense and easier deployment. Consequently, mesh networks enable a number of new application scenarios, including community wireless networking to provide affordable Internet access especially beneficial for low-income neighborhoods and scarcely populated areas.

Wireless mesh access points can have many multiple radio interfaces. Every interface uses an available channel (IEEE 802.11b/g standards have 3 non-overlapping channels in 2.4GHz band; original IEEE 802.11a standard have 12 non-overlapping channels in 5GHz band). Wireless mesh access points can communicate with many neighbor points at the same time. However, for propagation characteristics of wireless signal, node can receive all signals in its reception range and can send data equally in all directions of it. It has a serious drawback, namely channel interference [1]. Interference arises whenever two or more nodes try to communication on the same channel within each other's reception range at the same time. It results in high packet loss rate, high delay and low network throughput. Channel assignment algorithms are proposed for solving this problem, e.g. [2,3]. The goals of channel assignment algorithm are minimizing the interference between communicating links of wireless network and ensuring the quality of communication.

This paper addresses the channel assignment problem and specifically investigates the dynamic assignment of channels in a wireless mesh network. We present a routing-based dynamic channel assignment algorithm called R-CA that employs multiple radio channels simultaneously by equipping each node with multiple interfaces. The algorithm flexibly selects channels for the mesh radios in order to minimize interference within the communications in wireless mesh network. The rest of paper is structured as follows. Section II outlines some channel assignment algorithms. Section III proposes the R-CA strategy. Section IV evaluates the performance of R-CA. Section V concludes the paper.

II. RELATED WORK

The original studies of channel assignment problems are conducted in the field of mobile ad hoc networks. Based on these studies, more theory and algorithms for wireless mesh networks are proposed.

Kodialam and Nandagopai [1] conclude three kinds of collision avoidance model (interference model) which are PCA (primary conflict avoidance), RCA (receiver conflict avoidance), and TRCA (transmitter-receiver conflict avoidance). The interference of TRCA is most serious. In TRCA, when a node sends or receives packets, all nodes in this reception range can not communicate.

Raniwala *et al* [2] describe a topology network, and translates the channel assignment problem of this network into a NP difficult multiple subset problem.

Marina *et al* [3] propose a CLICA. It translates the channel assignment problem into coloring problem. In CLICA, the authors create a weighted conflict graph according to the protocol model and use this graph to show status of conflict channels. CLICA minimizes the maximum interference of each link, but it reduces the expansibility of channel assignment.

Subramanian *et al* [4] propose that the channel assignment problem is minimizing the maximum network interference when the number of radio frequency is limited.

*Corresponding author.

It is translated into conflict graph. The algorithm uses Tabu search technique to color the nodes of the conflict graph. Because it may make network shock, however, it can only be used in static network.

Michelle *et al* [5,6] propose that nodes use common channels to send and receive route information and control information. However, more network load and more routing nodes will lead to higher probability of data conflict.

The algorithms of [7] and [8] are based on packets. They may reduce throughput and increase delay time because of frequent channel switching.

Algorithms based on flows have been proposed in [9] and [10]. This kind of algorithms needs a complex iterative process, and it will make switching channel frequently when network traffic is changed.

Raniwala and Chiueh [11] propose a centralized interference-aware channel allocation BFS-CA. The algorithm is essentially a greedy algorithm. The authors consider data flows will appear interrupt when communicating nodes switch channel. These flows use default channels to transfer data during this time for avoiding loss. However, the algorithm needs every node use a channel to obtain the number interference radio. As a result, these channels are wasted.

Ramachandran *et al* [12] propose a hyacinth structure. It takes into account the stability of the wireless mesh network topology and tree aggregation traffic. It uses gateway as the root of the logical tree topology. The algorithm includes a load-balancing routing algorithm, an adaptive traffic load channel assignment algorithm and fault recovery mechanism. It also summarizes the purpose of channel assignment, how to solve the dependence problems of channel assignment and the principles of distribution channels.

### III. ROUTING-BASED DYNAMIC CHANNEL ASSIGNMENT ALGORITHM

#### A. Problem Description

For a successful allocation, it must satisfy the following conditions:

1) Consider the channel assignment problem on a static wireless network of $N$ nodes, each of which has $K$ interfaces. Let $C$ denote the number of available orthogonal channels, $C > 1$ and $C >= K$.

2) Given an assignment of channels to a communication, nodes $i$ and $j$ can communicate if and only if they share a common channel and they are within communication range of each other. If these conditions are satisfied, we say that a link exists between node $i$ and $j$. This is one of the goals of the channel assignment which is to ensure network connectivity.

3) It can avoid the interference among two or more reception ranges as much as possible. This is another goal of channel assignment which is ensuring the communication quality.

4) It can be easily used in the wireless mesh network and can adapt with respect to the variable transmitting qualities of wireless links.

To solve the dynamic channel assignment problems, a routing-based channel assignment algorithm called R-CA is proposed. The channels assigned on the links are dynamic changed with the next-hop routing direction.

#### B. Details of Algorithm

The details of algorithm will be described in this sub-section. R-CA needs additional information and it can assign proper channels to links based on the information.

R-CA needs every node to generate and maintain a table, and the main data is shown in Table 1. It includes the channel expected to occupy by communication ($c_{pre}$) which is initialized to 0, the channel occupied by communication ($c_{cur}$) which is initialized to 0, the expected completion time for data transmission ($t_{pre}$) which is initialized to 0, and waiting node address. $c_{pre}$ is used to denote the channel the node will use. $c_{cur}$ is channel the node is using now. $t_{pre}$ is celebrated by node and used in response management algorithm to control response information.

**Table 1. Channel information**

| Channel expected to occupy | $c_{pre}$ |
|---|---|
| Channel occupied | $c_{cur}$ |
| Expected completion time for data transmission | $t_{pre}$ |

The algorithm of R-CA includes two sections: distribution algorithm and response management algorithm. Distribution algorithm uses the result of response management algorithm.

**Algorithm 1**: Distribution algorithm
1:   do {
2:     if (not exist a channel $c_{pre}$ that is non-interrupting with $c_{cur}$)
3:       {wait for an available channel }
4:     else
5:       {while ($c_{pre}$ is available)
6:         {node update its $c_{cur}$;
7:          $c_{pre}$ =$c_{cur}$;
8:          route to next hop node with channel $c_{pre}$; }
9:         node calculates $t_{pre}$;
10:        if (no available channel)
11:          {use algorithm2 to decide the next step;
12:           if (respond routing to another node)
13:             {notice last hop to route to another node; }
14:           if (respond waiting)
15:             {store address;
16:              notice all hops behind to wait for available channel;
17:              notice next hop to store its address; }
18:          }
19:        else routing; }
20:  } while (find out the destination node and complete channel assignment)

Algorithm 1 given above is the main algorithm. It allocates channels when node routes packets and forwards frames to next hop. When there is no available channel to use,

Algorithm 1 uses Algorithm 2 to decide which channel should be allocated.

**Algorithm 2**: Response management algorithm
1:   receive routing packet;
2:   if (channel is unavailable)
3:     {compare receive $t_{pre}$(named $t_{pre}$' here) with its own $t_{pre}$;
4:     if ($t_{pre}$'<$t_{pre}$)
5:       {respond with information that route to another link;}
6:     else {respond with information that wait for free channel;
7:         $t_{pre}$ = $t_{pre}$ + $t_{pre}$';
8:         notice its address to next hop node;}
9:   }
10: else
11:   {respond with information that channel is available and keep on routing;}

When a node A routes a packet, its broadcast packet carries $c_{pre}$ besides routing information. The nodes around it receive the broadcast packet and update $c_{cur}$ if channel $c_{cur}$ is available. If other nodes route to one of these nodes, this node will decide the next step itself and do not affect node A. Our algorithm uses this method to avoid channel interrupting between adjacent nodes.

Algorithm 2 deals with response information. It is used when a node needs to change routing paths or wait for free channels.

In response management algorithm, there may be several nodes which are waiting for one node called *waited-node* to give up the channel at the same time. Then, when waited-node gives up the channel, which node first uses this node and this channel is an important problem. For this question, we define a queue to store the addresses of nodes that wait for channel (this kind of nodes is called *waiting-node*). Waited-node uses queue to store addresses of the waiting-nodes. After the waited-node gives up the channel, the first node in the queue can use this channel to route to waited-node. When the queue is full, waited-node refuses to store addresses of other nodes and notice the node which wants to store the address. The length of the queue can not be too large, which will cause the node to wait too long and impair the performance of network.

*C. Simple Examples*

This sub-section uses some examples to explain this algorithm. In the R-CA, source node can allocate channel according to the status of near nodes until finding the destination node. Channel is used when nodes need to communicate only and it will be released after communication. It ensures that nodes have enough channels to use. Three cases will appear with the algorithm. 1) At least one available channel can be used. Nodes use it directly. 2) No available channel can be used. Node routes through another node. 3) No available channel can be used. Node waits for a free channel. Next, there are some examples to show the whole process of the algorithm.

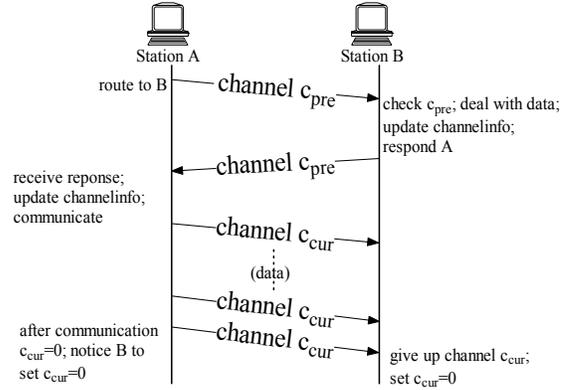

**Figure 1. Channel assignment diagram**

Fig.1 shows a simple channel assignment process of two nodes. After B receives channel information of A, B calculates $t_{pre}$ and update channelinfo. And then, B checks its $c_{cur}$ and received $c_{pre}$. If channel $c_{cur}$ and $c_{pre}$ are not interrupting, B adds $c_{pre}$ to its $c_{cur}$ and allocates channel $c_{cur}$ between A and B. If they are interferential, B compares received $t_{pre}$ (named $t_{pre}$' here) and its own $t_{pre}$. And then, respond waiting when $t_{pre}$'<$t_{pre}$. Otherwise, the response is routing through another node.

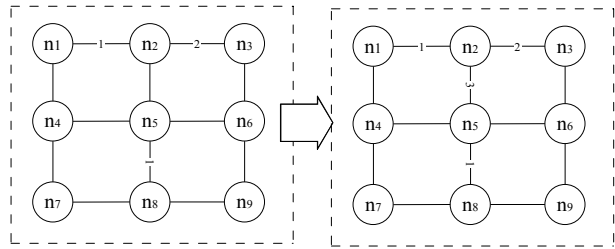

**Figure 2. Channel assignment of Case 1**

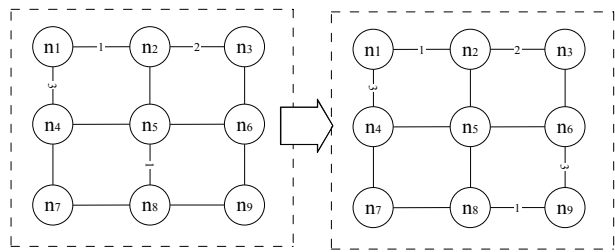

**Figure 3. Channel assignment of Case 2**

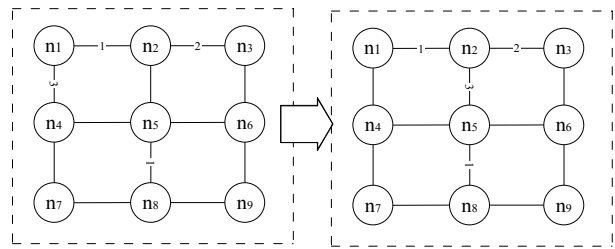

**Figure 4. Channel diagram of Case 3**

Figs. 2-4 are three simple examples to explain respectively Case 1, Case 2 and Case 3. In Figure 2, it shows that source node can route to destination node with non-interfering channel. Figure 3 shows the process that node routes through another node. Figure 4 shows the process of waiting for available channel. In three figures, it is supposed that there is no interference between different channels. Each node can affect adjacent nodes of it.

In Fig. 2, $n_2$ and $n_3$ is communicating with channel 2. Node $n_2$ update its channel information and broadcasts the information to its near nodes. Node $n_5$ changes value of its $c_{cur}$ to 2 after receiving the broadcast packet. When $n_5$ receives channel information of $n_8$, it checks its $c_{cur}$. Its $c_{cur}$ is not equal to 1. Nodes $n_5$ and $n_8$ can use channel 1 to communicate. Then, $n_5$ checks its $c_{cur}$. Channel 3 is available. Consequently, $n_4$ can route to $n_2$ directly. In Fig. 3, $n_2$ and $n_4$ notice $n_5$ that channel 1, 2 and 3 are not available. $n_5$ changes its $c_{cur}$. $n_8$ wants to use channel 1 to connect to $n_5$. $n_5$ finds that this channel is not available and goes to Algorithm 2. By computing time cost, $n_5$ notices $n_8$ to connect to another node.

In Fig. 4, $n_2$ and $n_4$ notice $n_5$ that channel 1, 2 and 3 are not available. $n_5$ changes its $c_{cur}$. $n_8$ wants to use channel 1 to connect to $n_5$. $n_5$ checks this channel is not available and goes to Algorithm 2. By comparing with time cost, $n_5$ notices $n_8$ to wait for a free channel. After the communication of $n_1$ and $n_4$, channel 3 is free, and then $n_5$ keeps on routing.

These examples show that R-CA can assign channels according to different scenarios. R-CA is a cross-layer and dynamic algorithm. Node can change channel and routing path according to channel information of its neighbors.

## IV. PERFORMANCE EVALUATION

The objective of the evaluation is to understand the behavior of the R-CA algorithm. In this section, we will describe the simulation environment and analyze the simulation results.

### A. Simulation Enviroment

R-CA is simulated in NS-2.34. The simulation environment is as follows. Every multi-radio node has 4 network interface cards. Simulation range is 1200×1200 in NS2. Each simulation time is 50s. The number of wireless node is 30. The data packet size is 512Byte. Every parameter in the channel information table is initialized to 0. We modify AODV routing protocol for carrying channel information when nodes route. In the new protocol, RREP packets include $c_{pre}$ and $t_{pre}$ besides routing information. RREQ packets add response information for the next hop. The length of waiting node address queue is 10 here. We use 10 data flows to simulate.

In the first scenario, the frequency of sending packet is variable from 5 to 25. The number of flows in the network is 4. Other parameters are fixed. We compare packet delivery rate of R-CA against a static channel assignment method and a single-radio method.

In the second scenario, all parameters are fixed except for the number of flows. This number change from 2 to 10. The frequency of sending packet is 20 packets per second. The average throughput of R-CA is compared against a static channel assignment method and a single-radio method.

### B. Simulation Results and Analysis

The simulation results are shown in Figures 5 and 6.

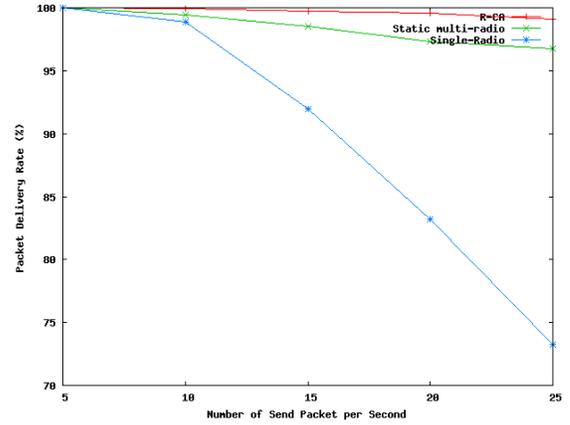

**Figure 5. Impact of the number of packets sent per second**

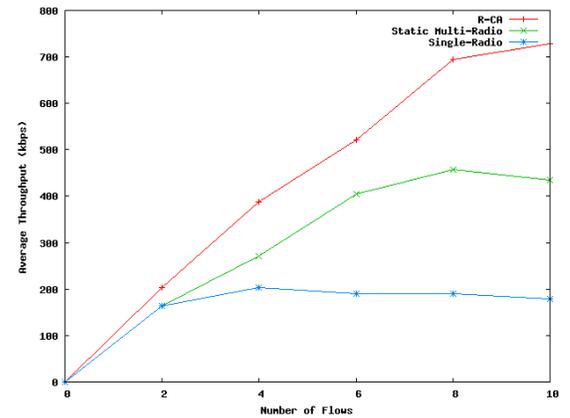

**Figure 6. Impact of the number of flows in the topology**

Fig. 5 plots the average packet delivery rate with different frequencies of sending packet in a single-radio mesh network, in a multi-radio mesh network with a static assignment and in a multi-radio mesh network with R-CA, respectively. At the beginning, three packet delivery rates are similar. When the frequency of sending packet is 10 packets per second, the packet delivery rate of single-radio begins to decrease faster. The packet delivery rates of static multi-radio and R-CA decrease insignificantly.

In single-radio solutions, each node switches channels dynamically and randomly selects the channels, while coordinating with neighbor nodes to ensure communication over a common channel for some periods. Such coordination, however, requires tight time synchronization among nodes. Slow switching between channels can reduce synchronization requirements and overheads, but increases end-to-end delay time and packet delivery rate. R-CA and static channel assignment algorithm do not switch channel when nodes receive or send packets. This leads to higher packet delivery rate.

Fig. 6 plots the average throughput of flows in a single-radio mesh network, in a multi-radio mesh network with a static assignment and in a multi-radio mesh network with R-CA, respectively. At the beginning, the average throughput of three cases is similar. This is because less flow yields less channel disturbances. With the increase of flows, the orthogonal channels become less relative. Because of the mode of communication of single-radio, its throughput is lowest. When the number of flows ranges from 8 to 10, the average throughput of static multi-radio assignment decreases, as well as the increment of R-CA. The reason behind is that each node using static multi-radio assignment has a fixed number of interfaces (in our simulations, because of four neighbors of each node, this number is 4), and every interface is defined as a channel. Too many flows lead to unfair loads in network. Packets are too many to transfer in some links and some links do not work. Nodes using R-CA can release the channels after transmission. Thus nodes have more usable resource. When the number of flows is more than 8, some nodes put addresses of nodes waiting for available channels into waiting node queue. This makes the increment decrease. As a result, the average throughput of R-CA is much closer to the optimal. It has much better capacity than static multi-radio and single-radio although the increment decreases when the number of flows becomes large.

In simulations, the results of channel assignment by using R-CA are the best. The performance of the network can be improved wit R-CA. For example, R-CA outperforms static multi-radio channel assignment algorithm about 17%, 57%, 67% in terms of average throughput when the number of flow is 2, 8 and 10, respectively.

## V. CONCLUSION

In this paper, some typical existing channel assignment algorithms have been analyzed and the principles of channel distribution have been formulated. Considering the problem of channel interference in multi-radio wireless mesh networks, a dynamic channel assignment algorithm called R-CA to find connected and low interference topologies has been developed. R-CA is a cross-layer algorithm including MAC layer and network layer, because channel information and allocation method can decide the next hop of routing. Simulation results demonstrate the effectiveness of the R-CA algorithm in exploiting channel assignment diversity for reducing interference with a small number of radios per node, and resulting in significant performance benefits in an 802.11-based multi-radio mesh network with single hop as well as multi-hop workloads.

As the number of the wireless nodes increases, more routing and channel information will require more computations. Nodes then need more time to decide the next hop. This time will constrain the performance of the algorithm. Our future work will focus on reducing the time cost and more comprehensive evaluation of the R-CA algorithm, including theoretical characterization and real-world performance evaluation in a multi-radio wireless mesh testbed.

ACKNOWLEDGMENT

This work is supported in part by Natural Science Foundation of China under grant No. 60903153 and the Fundamental Research Funds for the Central Universities.